\documentclass[fleqn,usenatbib]{mnras}

\usepackage{newtxtext,newtxmath}
\usepackage[T1]{fontenc}

\DeclareRobustCommand{\VAN}[3]{#2}
\let\VANthebibliography\thebibliography
\def\thebibliography{\DeclareRobustCommand{\VAN}[3]{##3}\VANthebibliography}

\usepackage{graphicx}	% Including figure files
\usepackage{amsmath}	% Advanced maths commands
\usepackage{multirow} % martians-yufan

\title[Martians see more PHAs than Earthlings]{MARTIANS (MARs2020, TIANwen and So on) would see more potentially hazardous asteroids than Earthlings}

\author[Y. F. Zhou et al.]{
Yufan Fane Zhou,$^{1,2}$\thanks{E-mail: yufanz@smail.nju.edu.cn}
Hailiang Li,$^{1,2}$
Zhiyuan Li$^{1,2,3}$\thanks{E-mail: lizy@nju.edu.cn}
and Liyong Zhou$^{1,2}$
\\
$^{1}$School of Astronomy and Space Science, Nanjing University, Nanjing 210023, China\\
$^{2}$Key Laboratory of Modern Astronomy and Astrophysics (Nanjing University), Ministry of Education, Nanjing 210023, China\\
$^{3}$Institute of Science and Technology for Deep Space Exploration, Suzhou Campus, Nanjing University, Suzhou 215163, China\\
}

\date{Accepted 2024 May 03. Received 2024 May 03; in original form 2024 January 30}

\pubyear{2024}

\begin{document}
\label{firstpage}
\pagerange{\pageref{firstpage}--\pageref{lastpage}}
\maketitle

\begin{abstract}
Potentially Hazardous Asteroids (PHAs) are a special subset of Near-Earth Objects (NEOs) that can come close to the Earth and are large enough to cause significant damage in the event of an impact. Observations and researches of Earth-PHAs have been underway for decades. Here, we extend the concept of PHAs to Mars and study the feasibility of detecting Mars-PHAs in the near future. We focus on PHAs that truly undergo close approaches with a planet (dubbed CAPHAs) and aim to compare the actual quantities of Earth-CAPHAs and Mars-CAPHAs by conducting numerical simulations incorporating the Yarkovsky effect, based on observed data of the main asteroid belt. The estimated number of Earth-CAPHAs and Mars-CAPHAs are 4675 and 16910, respectively. The occurrence frequency of Mars-CAPHAs is about 52 per year, which is 2.6 times that of Earth-CAPHAs, indicating significant potential for future Mars-based observations. Furthermore, a few Mars-CAPHAs are predicted to be observable even from Earth around the time of next Mars opposition in 2025. 

\end{abstract}


%%%%%%%%%%%%%%%%%%%%%%%%%%%%%%%%%%%%%%%%%%%%%%%%%%

%%%%%%%%%%%%%%%%% BODY OF PAPER %%%%%%%%%%%%%%%%%%
\section{Introduction}
\label{sec:introduction}
Near-Earth objects (NEOs) are small bodies whose perihelion distances $q\leq 1.3$\,au and aphelion distances $Q\geq 0.983$\,au, including near-Earth asteroids (NEAs) and near-Earth comet nuclei (NECs). Potentially hazardous asteroids (PHAs), a special subset of NEAs, are defined as having a minimum orbit intersection distance (MOID) to Earth orbit of less than 0.05\,au and absolute magnitude $H\leq 22$\,mag (i.e., diameter $D\geq 140$\,m). PHAs are more likely to collide with Earth than other NEAs, and if such a collision were to occur, their size is large enough to cause destructive effects \citep[for a review see][]{perna2013}. However, PHAs are not necessarily highly `dangerous' because orbital proximity does not imply proximity in actual distance (e.g. the case of Earth Trojan asteroids). Therefore, only those PHAs that truly undergo close approaches (hereafter called CAPHAs) to the Earth are the primary objects of human monitoring\footnote{ https://cneos.jpl.nasa.gov/ca}, in which case the asteroid has a minimum distance $r_{\rm min} <$ 0.05\,au from Earth.

NEAs have been objects of human attention for decades, motivated by concerns for both scientific interests and our own safety. On one hand, such objects could have delivered water and organic-rich materials to the early Earth, which is crucial for our understanding about the origin of terrestrial life as well as the formation and early evolution of the Solar System \citep[e.g.][]{marty2012,alexander2012,altwegg2015}. On the other hand, potential impact events can pose significant threats to the Earth. The most famous one is undoubtedly the so-called `K-T' event leading to the Cretaceous/Tertiary extinction that occurred about 65 million years ago \citep{kyte1998,bottke2007}. The Tungus Explosion in 1908 \citep{ben1975} and the Chelyabinsk meteor falling in 2013 \citep{brown2013} are also well-known examples.

According to the Small-Body Database\footnote{ https://ssd.jpl.nasa.gov/tools/sbdb\_query.html} (SBDB) of Jet Propulsion Laboratory (JPL), as of June 26, 2023, over 30000 NEAs have been observed. The dynamical lifetime of NEAs is typically between $10^6$ and $10^8$ years \citep{morbidelli1998}, so asteroids must be continually transported to maintain the population we observe around Earth today \citep{mainzer2011}. Studies by \citet{wetherill1988}, \citet{rabinowitz1997a,rabinowitz1997b}, \citet{bottke2002} suggest that NEAs are probably sourced from unstable regions in the main asteroid belt, such as the 3:1 mean motion resonance with Jupiter \citep{wisdom1983,moons1996} and the $\nu_6$ secular resonance \citep{morbidelli1994}, which manifest as `gaps' in the orbital distribution of asteroids. However, precisely because such regions are unstable, the number of asteroids in these locations is relatively low. Therefore, a mechanism that can push nearby asteroids into these resonance regions must be considered: the Yarkovsky effect.

The Yarkovsky effect \citep{burns1979,bottke2006} was first discovered by the engineer Ivan Osipovich Yarkovsky in 1901. About a century later, it was shown to have an important influence on the motion of artificial satellites \citep{rubincam1987,rubincam1995} and small bodies \citep{farinella1998}. The Yarkovsky effect is caused by the asymmetric thermal emission (with respect to the subsolar point) from a rotating, atmosphereless body with finite thermal inertia, and it manifests itself as a recoil force. Radiation from the Sun heats up asteroids, with the side facing the Sun having the highest temperature. After a short period of time, the asteroid will re-radiate its thermal energy and generate a net radiation imbalance. As the asteroid is rotating, its hottest side may no longer be perfectly aligned with the Sun. Therefore, the recoil force has a transverse component, which will accelerate or decelerate the asteroid's orbital motion, thus change the semi-major axis of the asteroid over a reasonable long time scale \citep{peterson1976,farinella1998,farinella1999}.

The above concepts are also applicable to asteroids near Mars. However, numerical simulations and observations of such objects are currently very limited \citep[e.g.][]{migliorini1998,michel2000,ali-lagoa2017}, and the focus has mainly been on Mars-crossing asteroids (MCAs), which are defined as asteroids that can cross the orbit of Mars. Here we focus on asteroids that truly close to Mars and propose the concept of Mars-CAPHAs. For Mars-CAPHAs, we set the distance criterion as $r_{\rm min} <$ 0.036\,au, instead of 0.05\,au, to take into account the difference in the Hill radius between Earth and Mars (choosing 0.05\,au for Mars is also acceptable and will result in more Mars-CAPHAs).

The study of Mars-CAPHAs will not be merely conceptual. Indeed, humanity possesses the capability to investigate them \textit{in situ}. Since the 1960s, over forty Mars exploration missions have been conducted. With the advancement of technology, missions have progressed from early attempts at flybys, such as Mariner-4 \citep{huntress2003}, to include orbiters, landers and rovers, such as 2001 Mars Odyssey \citep{saunders2004}, Mars Express \citep{nielsen2004} and Zhu Rong \citep{li2021}. In the future, projects such as Mars~2020 from the USA (Perseverance has already arrived on Mars) \citep{farley2020}, Tianwen-3 from China, ExoMars from Europe, among others, will utilize orbiters and rovers to carry out more in-depth explorations of Mars. Observing and imaging Mars-CAPHAs with Mars-based instruments, although challenging, may bring us new insights into the Martian environment.

Therefore, numerical simulations about Mars-CAPHAs are timely and necessary to assess their observational prospects. We conduct an \textit{N}-body simulation incorporating the Yarkovsky effect for such a study. The methods are presented in section~\ref{sec:methods}. We introduce the software and the implementation of the Yarkovsky effect. Information about the data used in this work and our simulation setup is also provided. In section~\ref{sec:results}, we estimate some observable properties of CAPHAs and address their implications. A brief summary is presented in section~\ref{sec:summary}.

\section{Data and Methods}
\label{sec:methods}
\subsection{\textit{Mercury6} with the Yarkovsky force}
The numerical simulation tool we use is the \textit{N}-body software package \emph{Mercury6} \citep{chambers1999}, which can integrate any star-planet system like our Solar System, with an additional Yarkovsky force subroutine. The hybrid integrator is utilized, which is basically a symplectic integrator switching to Bulirsch-Stoer (an algorithm with adaptive timesteps) only when planetary close encounters occur. We incorporated the Yarkovsky force into the subroutine \emph{mfo\_user} using the same approach as \citet{zhou2019}.

The Yarkovsky effect results in a recoil force on asteroids, and the precise calculation of this force is highly complex and dependent on lots of physical parameters, many of which have not yet been accurately measured. In some studies \citep[e.g.][]{vokrouhlicky1998,vokrouhlicky1999}, the Yarkovsky effect is simply described as an equivalent drift rate of the semi-major axis ${\rm d}a/{\rm d}t$ of an asteroid. \citet{xu2020} calculated that an asteroid located at $a=2.9$\,au with diameter $D=1$\,km has a drift rate of ${\rm d}a/{\rm d}t = 0.256$\,au\,Gyr$^{-1}$, assuming rotation period $P=8$\,h, albedo $p=0.13$, thermal conductivity $K=0.005$\,W\,m$^{-1}$\,K$^{-1}$, specific heat capacity $C=680$\,J\,kg$^{-1}$\,K$^{-1}$, surface density $\rho_{\rm s}=1.5$\,g\,cm$^{-3}$, bulk density $\rho=2.5$\,g\,cm$^{-3}$ and spinning obliquity $\gamma=0$\,deg \citep{vokrouhlick2006}. Although \citet{xu2020b} indicates that the relationship between the drift rate and semi-major axis may be complex, ${\rm d}a/{\rm d}t$ can be considered inversely proportional to $a^2$ within the range of the main belt \citep[e.g.][]{nesvorny2023}. Additionally, the drift rate is inversely proportional to the diameter $D$.

According to \citet{xu2020}, for an asteroid with diameter $D=1$\,km in the 7:3 mean motion resonance (2.958\,au), we set its drift rate to ${\rm d}a/{\rm d}t = 0.25$\,au\,Gyr$^{-1}$. Then following the same rules as \citet{nesvorny2023}, the formula for drift rate used in our work is
\begin{equation}
    \frac{{\rm d}a}{{\rm d}t}=0.25\,\frac{\rm au}{\rm Gyr}\times \left(\frac{1\,{\rm km}}{D}\right)\left(\frac{2.958\,{\rm au}}{a}\right)^{2}.
	\label{eq:dadt}
\end{equation}

\subsection{Asteroid data sampling and grouping}
Earth-CAPHAs and Mars-CAPHAs mainly originate from the main belt \citep[e.g.][]{rabinowitz1997a,rabinowitz1997b}, which has been stable for a considerable period of time. During this time, the escape of asteroids from the main belt does not significantly change its distribution. Based on this assumption, we use observed data of main belt asteroids as the initial conditions for our simulations. From the SBDB of JPL, as of June 26, 2023, we downloaded the data containing a total of 1,223,263 asteroids.

\begin{figure}
    \includegraphics[width=\columnwidth]{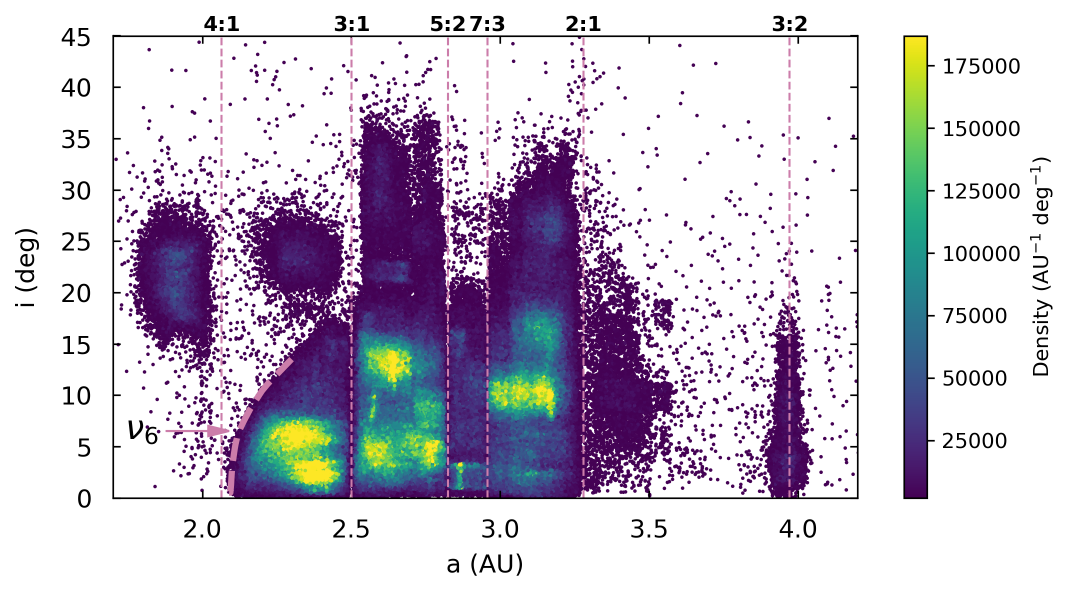}
    \caption{Scatter plot of inclination $i$ versus semi-major axis $a$ of the main belt (1,223,263 asteroids). The color reflects the number density of the scatter points. Several mean motion resonances and the $\nu_6$ secular resonance are marked with reddish purple dashed lines.}
    \label{fig:1}
\end{figure}

Fig.~\ref{fig:1} shows the scatter plot of inclination $i$ versus semi-major axis $a$ of the main belt based on the 1,223,263 asteroids. Several strong mean motion resonances and the $\nu_6$ secular resonance are marked. After removing those with large measurement errors ($\sigma_{\rm a}\geq 0.0001$\,au), 1,144,461 asteroids remain. For a feasible computational cost, we randomly sampled 1\% (11445 asteroids) as our `sample data'.

We divide the 11445 asteroids into two categories named `near-gap' asteroids (709) and `far-gap' asteroids (10736) (specific ranges of `near-gap' and `far-gap' regions can be found in column~3 of Table~\ref{tab:setting_result}). Since different mean motion resonances have different widths, sizes of our six `near-gap' regions should be set differently. We set the range of each `near-gap' region to be significantly wider than its resonance width and to contain a sufficient number of asteroids. For different mean motion resonances, the lower its order is, the larger width we assign to it. It is worth mentioning that the 4:1 resonance is very close to the $\nu_6$ resonance, so we set a relatively large width for it and extend the right edge to 2.200\,au \citep{wetherill1985} to embrace the $\nu_6$ resonance \citep[e.g.][]{scholl1991}. For other relatively strong resonances not shown in Fig.~\ref{fig:1}, such as the 5:1 resonance (1.780\,au), even setting a particularly large width does not include more than 10 asteroids in our sample, so we do not consider them while studying `near-gap' asteroids.

\subsection{Simulation settings}
\label{subsec:setting}
The dynamical model used in our simulations consists of the Sun, eight planets from Mercury to Neptune and (massless) asteroids. The Earth-Moon barycentre is adopted instead of the separate Earth and Moon \citep[e.g.][]{dvorak2012,zhou2019}. The initial data of eight planets at epoch JD 2457724.5 are obtained from the JPL HORIZONS system \citep{giorgini1996}. Asteroids with different epochs from planets are integrated to the same epoch before the main integration begins. The initial timestep and accuracy parameter we set are 8 days and 1e-12, respectively. Our timestep is a bit longer compared to other work \citep{granvik2018}, but here we focus on the statistical level results rather than high precision orbital integration of any specific body. In our simulations, an asteroid will no longer be tracked after being ejected from the Solar System (defined as being at a distance greater than 100\,au from the Sun).

For `near-gap' asteroids, we perform simulations with the Yarkovsky effect for 0.1\,Gyr, which is sufficiently long for the effect to become evident. For `far-gap' asteroids, we neglect the Yarkovsky effect (i.e. only with gravity) and run the simulations for 1\,Myr only, a simplification in order to save computational effort. \citet{granvik2017} showed that the Yarkovsky effect can deliver 100-metre-diameter asteroids from most regions in the main belt to $q=1.3$\,au within tens of millions of years, but such a delivery predicts too many NEAs compared to observations. On the other hand, the combined Yarkovsky and YORP \citep{bottke2006} model yields results lower than the observed values, possibly due to the need for improvement in the canonical YORP model \citep{granvik2017}. Therefore, currently there are some issues with the drift timescale of main belt asteroids caused by the Yarkovsky (and YORP) effect. Nevertheless, for `near-gap' asteroids, the drift timescale does not matter much, as they are initially close to strong resonances, hence considering only the Yarkovsky effect can yield results consistent with observations \citep[e.g.][]{xu2020b}. Moreover, since we are mostly concerned with asteroids with sizes greater than 100-metre, which would take an even longer time to excite, `far-gap' asteroids should contribute only a minor fraction of the resultant CAPHAs. Thus, our separate treatment for the `near-gap' and `far-gap' asteroids is justified.

When calculating the Yarkovsky drift rate of an asteroid, its semi-major axis $a$ and diameter $D$ are required. Here, for a `near-gap' region, $a$ of all asteroids within it are approximated by the value of the center of the region. However, $D$ of most asteroids are still unknown. Additionally, observations of the main belt are incomplete: asteroids with small $D$ (i.e., large absolute magnitude $H$) must have been missed so far, and such selection effect becomes more severe for more distant regions. Therefore, we generate multiple sets of cloned asteroids for each `near-gap' region, assigning them different $H$ values in the simulations, in order to prepare for subsequent correction for the selection effect. $H$ and $D$ follow the empirical conversion formula:
\begin{equation}
    D=\frac{1329\,{\rm km}}{\sqrt{p}}\times 10^{-0.2H}.
	\label{eq:hd}
\end{equation}
Here albedo $p$ is set to 0.14 \citep{mainzer2011b}, which is close to the value used in the calculation of the drift rate by \citet{xu2020}.

The absolute magnitude range we consider is from 15\,mag (the number of asteroids with $H<15$\,mag is small) to 22\,mag (the upper limit of CAPHA's definition). We divide the range into 7 intervals, with each interval spanning 1\,mag. For each interval, we approximate the entire range using the central magnitude value.

The specific settings for our simulations are shown in columns~1-3 of Table~\ref{tab:setting_result}, where six `near-gap' regions are designated by A-F and the `far-gap' region is designated by G. For each `near-gap' region, a total of 8 simulations are conducted. Simulation~0 serves as the control simulation without the Yarkovsky effect, while simulations~1-7 correspond to the $H$ value ranging from 15.5 to 21.5.

Furthermore, in reality, depending on the direction of rotation of an asteroid, the Yarkovsky effect can either increase or decrease its semi-major axis. However, rotation directions of most asteroids are unknown. Therefore, for each `near-gap' region, we set the direction of the Yarkovsky force such that the asteroid is drifted toward the center of the gap. Such a treatment artificially raises the number of CAPHAs by a factor of two, which is subsequently corrected in section~\ref{sec:results}. A related caveat is that this treatment prevents asteroids from drifting into more distant resonances in the opposite direction, but the fractional contribution of CAPHAs through this channel is expected to be negligible.

\begin{table*}
	\centering
	\caption{Simulation ID, Yarkovsky drift rate, region information (including the radial range, corresponding resonance and number of asteroids in our sample), number of CAPHAs ($N$), difference caused by the Yarkovsky effect ($\Delta N$), reasonable number ($N^{*}=N_{\rm 0}+\frac{1}{2}\Delta N$), average active coefficient ($\bar \epsilon$), supplement coefficient ($\eta$) and currently observable number ($N_{\rm obs}$) of our simulations. Series~A-F belong to `near-gap' asteroids and correspond to six mean motion resonances (4:1, 3:1, 5:2, 7:3, 2:1, 3:2). Simulation~0 in each series corresponds to the control group, while simulations~1-7 correspond to the cases where the absolute magnitudes (diameters) are 15.5\,mag (2.82\,km), 16.5\,mag (1.78\,km), 17.5\,mag (1.12\,km), 18.5\,mag (0.71\,km), 19.5\,mag (0.45\,km), 20.5\,mag (0.28\,km) and 21.5\,mag (0.18\,km), respectively. Series~G corresponds to `far-gap' asteroids. Zero drift rate means no Yarkovsky effect. Except for series~G with a duration of 0.001\,Gyr, all other simulations have a duration of 0.1\,Gyr. `EC' and `MC' are abbreviations for Earth-CAPHA and Mars-CAPHA. The two $\eta$ values for series~G correspond to the values of Earth-CAPHAs and Mars-CAPHAs, respectively. The range (au) of series F (3:2 resonance) is $3.972\pm 0.050$, whose result is not shown as it contributes insignificantly to the population of CAPHAs.}
	\label{tab:setting_result}
	\begin{tabular}{cccccccccccccc}
		\hline
		\multirow{2}*{ID} & drift rate & region & \multirow{2}*{$N_{\rm EC}$} & \multirow{2}*{$N_{\rm MC}$} & \multirow{2}*{$\Delta N_{\rm EC}$} & \multirow{2}*{$\Delta N_{\rm MC}$} & \multirow{2}*{$N_{\rm EC}^{*}$} & \multirow{2}*{$N_{\rm MC}^{*}$} & \multirow{2}*{$\bar \epsilon_{\rm EC}$} & \multirow{2}*{$\bar \epsilon_{\rm MC}$} & \multirow{2}*{$\eta$} & \multirow{2}*{$N_{\rm EC,obs}$} & \multirow{2}*{$N_{\rm MC,obs}$} \\
            ~ & (au\,Gyr$^{-1}$) & information & ~ & ~ & ~ & ~ & ~ & ~ & ~ & ~ & ~ & ~ \\
            (1) & (2) & (3) & (4) & (5) & (6) & (7) & (8) & (9) & (10) & (11) & (12) & (13) & (14) \\
		\hline
        A0 & 0 & range (au): & 61 & 87 & - & - & - & - & - & - & - & - & -\\
        A1 & 0.18 & $2.065^{+0.135}_{-0.050}$ & 72 & 125 & 11 & 38 & 66.5 & 106.0 & 0.061 & 0.480 & 0.051 & 0.206 & 2.584\\
        A2 & 0.29 & ~ & 96 & 148 & 35 & 61 & 78.5 & 117.5 & 0.049 & 0.446 & 0.115 & 0.445 & 6.030\\
        A3 & 0.46 & resonance: & 123 & 170 & 62 & 83 & 92.0 & 128.5 & 0.074 & 0.397 & 0.261 & 1.771 & 13.308\\
        A4 & 0.72 & 4:1 and $\nu_6$ & 164 & 189 & 103 & 102 & 112.5 & 138.0 & 0.087 & 0.354 & 0.236 & 2.318 & 11.512\\
        A5 & 1.13 & ~ & 191 & 198 & 130 & 111 & 126.0 & 142.5 & 0.087 & 0.326 & 0.355 & 3.904 & 16.512\\
        A6 & 1.82 & population: & 198 & 198 & 137 & 111 & 129.5 & 142.5 & 0.118 & 0.301 & 0.535 & 8.155 & 22.964\\
        A7 & 2.83 & 198 & 198 & 198 & 137 & 111 & 129.5 & 142.5 & 0.108 & 0.234 & 0.806 & 11.267 & 26.829\\
        \hline
        B0 & 0 & range (au): & 32 & 32 & - & - & - & - & - & - & - & - & -\\
        B1 & 0.12 & $2.502\pm 0.025$ & 58 & 60 & 26 & 28 & 45.0 & 46.0 & 0.021 & 0.038 & 0.079 & 0.074 & 0.137\\
        B2 & 0.20 & ~ & 73 & 73 & 41 & 41 & 52.5 & 52.5 & 0.013 & 0.019 & 0.198 & 0.136 & 0.193\\
        B3 & 0.31 & resonance: & 89 & 92 & 57 & 60 & 60.5 & 62.0 & 0.013 & 0.039 & 0.497 & 0.391 & 1.206\\
        B4 & 0.49 & 3:1 & 89 & 90 & 57 & 58 & 60.5 & 61.0 & 0.024 & 0.043 & 0.483 & 0.695 & 1.267\\
        B5 & 0.78 & ~ & 96 & 99 & 64 & 67 & 64.0 & 65.5 & 0.026 & 0.025 & 0.765 & 1.249 & 1.238\\
        B6 & 1.25 & population: & 99 & 99 & 67 & 67 & 65.5 & 65.5 & 0.033 & 0.040 & 1.212 & 2.611 & 3.191\\
        B7 & 1.94 & 122 & 102 & 101 & 70 & 69 & 67.0 & 66.5 & 0.021 & 0.041 & 1.919 & 2.688 & 5.272\\
        \hline
        C0 & 0 & range (au): & 27 & 27 & - & - & - & - & - & - & - & - & -\\
        C1 & 0.10 & $2.825\pm 0.020$ & 44 & 44 & 17 & 17 & 35.5 & 35.5 & 0.004 & 0.007 & 0.120 & 0.017 & 0.028\\
        C2 & 0.15 & ~ & 50 & 50 & 23 & 23 & 38.5 & 38.5 & 0.010 & 0.013 & 0.322 & 0.124 & 0.159\\
        C3 & 0.24 & resonance: & 70 & 70 & 43 & 43 & 48.5 & 48.5 & 0.005 & 0.008 & 0.868 & 0.215 & 0.324\\
        C4 & 0.38 & 5:2 & 86 & 87 & 59 & 60 & 56.5 & 57.0 & 0.008 & 0.007 & 0.885 & 0.385 & 0.343\\
        C5 & 0.60 & ~ & 96 & 97 & 69 & 70 & 61.5 & 62.0 & 0.004 & 0.015 & 1.453 & 0.357 & 1.324\\
        C6 & 0.96 & population: & 91 & 92 & 64 & 65 & 59.0 & 59.5 & 0.004 & 0.007 & 2.384 & 0.535 & 0.936\\
        C7 & 1.50 & 132 & 99 & 98 & 72 & 71 & 63.0 & 62.5 & 0.004 & 0.007 & 3.913 & 0.986 & 1.810\\
        \hline
        D0 & 0 & range (au): & 5 & 8 & - & - & - & - & - & - & - & - & -\\
        D1 & 0.09 & $2.958\pm 0.015$ & 20 & 32 & 15 & 24 & 12.5 & 20.0 & 0.002 & 0.006 & 0.120 & 0.003 & 0.015\\
        D2 & 0.14 & ~ & 38 & 50 & 33 & 42 & 21.5 & 29.0 & 0.002 & 0.011 & 0.322 & 0.015 & 0.101\\
        D3 & 0.22 & resonance: & 42 & 59 & 37 & 51 & 23.5 & 33.5 & 0.010 & 0.013 & 0.868 & 0.202 & 0.366\\
        D4 & 0.35 & 7:3 & 49 & 72 & 44 & 64 & 27.0 & 40.0 & 0.001 & 0.007 & 0.885 & 0.024 & 0.244\\
        D5 & 0.56 & ~ & 41 & 68 & 36 & 60 & 23.0 & 38.0 & 0.005 & 0.011 & 1.453 & 0.164 & 0.624\\
        D6 & 0.89 & population: & 52 & 69 & 47 & 61 & 28.5 & 38.5 & 0.002 & 0.009 & 2.384 & 0.156 & 0.799\\
        D7 & 1.39 & 125 & 51 & 70 & 46 & 62 & 28.0 & 39.0 & 0.014 & 0.018 & 3.913 & 1.523 & 2.778\\
        \hline
        E0 & 0 & range (au): & 7 & 10 & - & - & - & - & - & - & - & - & -\\
        E1 & 0.07 & $3.279\pm 0.050$ & 19 & 26 & 12 & 16 & 13.0 & 18.0 & 0.003 & 0.011 & 0.172 & 0.007 & 0.034\\
        E2 & 0.11 & ~ & 31 & 35 & 24 & 25 & 19.0 & 22.5 & 0.008 & 0.022 & 0.470 & 0.075 & 0.234\\
        E3 & 0.18 & resonance: & 36 & 48 & 29 & 38 & 21.5 & 29.0 & 0.010 & 0.021 & 1.284 & 0.262 & 0.782\\
        E4 & 0.28 & 2:1 & 51 & 54 & 44 & 44 & 29.0 & 32.0 & 0.007 & 0.024 & 1.322 & 0.284 & 1.015\\
        E5 & 0.44 & ~ & 60 & 67 & 53 & 57 & 33.5 & 38.5 & 0.010 & 0.020 & 2.185 & 0.739 & 1.666\\
        E6 & 0.71 & population: & 58 & 67 & 51 & 57 & 32.5 & 38.5 & 0.007 & 0.028 & 3.611 & 0.868 & 3.893\\
        E7 & 1.11 & 77 & 59 & 66 & 52 & 56 & 33.0 & 38.0 & 0.014 & 0.027 & 5.968 & 2.659 & 6.214\\
		\hline
        G & 0 & population: 10736 & 3 & 152 & - & - & 3 & 152 & 0.244 & 0.353 & 1.7\&0.6 & 1.233 & 33.161\\
		\hline
        total & - & - & - & - & - & - & - & - & - & - & - & 46.75 & 169.10\\
        \hline
	\end{tabular}
\end{table*}

\section{Results and Implications}
\label{sec:results}
\subsection{Quantities of CAPHAs}
\label{sec:results_quantity}
The direct outputs from our simulations are presented in columns~4-5 of Table~\ref{tab:setting_result}, which are subject to further processing. First, for simulations~1-7, we calculate the increments $\Delta N$ relative to simulation~0. Considering that our setting for the direction of the Yarkovsky force would result in the increments being doubling the actual values, we correct the results of simulations~1-7 to $N^{*}=N_{\rm 0}+\frac{1}{2}\Delta N$, where $N_{\rm 0}$ is the number of CAPHAs predicted by simulation~0.

Second, we introduce the active coefficient
\begin{equation}
    \epsilon=\frac{t_{\rm act}}{T_{\rm sim}},
	\label{eq:epsilon}
\end{equation}
where $T_{\rm sim}$ is the simulation duration and $t_{\rm act}$ is the active lifetime. The active lifetime is the time difference between an asteroid's first and last occurrence as a CAPHA; for asteroids with just a single occurrence as a CAPHA, the active lifetime is set to be 5 years, which corresponds to the orbital period for $a=3$\,au (the center of our studied range). We introduce this coefficient because the quantities obtained from our simulations do not represent the current observable numbers. For example, if one CAPHA, whose active lifetime is 1\,Myr, is generated in a simulation spanning 100\,Myr, we consider its current survival probability to be 0.01, so its contribution to the total count is not 1 but rather 0.01. Average $\epsilon$ values for different simulations are presented in columns~10-11 of Table~\ref{tab:setting_result}.

Third, we introduce the supplement coefficient ($\eta$) to correct the selection effect of the main belt. Considering the different completeness at different distances, we partition the main belt into five zones named $\romannumeral1$, $\romannumeral2$, $\romannumeral3$, $\romannumeral4$ and $\romannumeral5$, using 2.3\,au, 2.7\,au, 3.1\,au and 3.5\,au as dividing points. Our `near-gap' simulations A, B, C\&D, E, F respectively correspond to zone $\romannumeral1$, $\romannumeral2$, $\romannumeral3$, $\romannumeral4$ and $\romannumeral5$. Assuming that the actual cumulative magnitude distribution of asteroids follows an exponential function \citep[e.g.][]{gladman2009}, we conduct fittings using asteroids with $15\leq H\leq H^{\prime}$ in each of the five zones. The values of $H^{\prime}$ are chosen based on the observed cumulative magnitude distribution and are respectively set to 18.0, 18.0, 17.3, 17.0 and 16.2. We obtain:
\begin{equation}
    lg[N(<H)]=\left\{
              \begin{aligned}
              \begin{array}{ll}
              0.36H-1.68, & \romannumeral1\,(a<2.3)\\
              0.40H-1.65, & \romannumeral2\,(2.3\leq a<2.7)\\
              0.43H-2.10, & \romannumeral3\,(2.7\leq a<3.1)\\
              0.44H-2.32, & \romannumeral4\,(3.1\leq a<3.5)\\
              0.41H-3.03, & \romannumeral5\,(a\geq 3.5)\\
              \end{array}
              \end{aligned}
              \right
              .
	\label{eq:exp}
\end{equation}
Furthermore, there is strong evidence showing that the size-frequency distribution of the main belt flattens for $H>18$ \citep[e.g.][]{gladman2009}, so we assume that the slopes of the five sub-equations are halved when $H$ reaches 18. Therefore, for each zone, we can deduce the ratio of the number of asteroids in 7 magnitude intervals (corresponding to simulations 1-7), and we define $q_{\rm xk}$ as the $k$-th value among the 7 ratios of zone $x$. For example, the ratio for zone~$\romannumeral4$ is 1.00~:~2.73~:~7.46~:~7.68~:~12.70~:~20.98~:~34.68, where the first value ($q_{\rm \romannumeral4 1}$, set to 1.00) corresponds to 15\,mag-16\,mag, and the last value ($q_{\rm \romannumeral4 7}$) corresponds to 21\,mag-22\,mag. However, due to the selection effect against faint objects, the observed distributions significantly differ from theoretical ones. Therefore, we introduce the supplement coefficient for zone $x$, magnitude interval $k$:
\begin{equation}
    \eta_{\rm xk}=\frac{n_{\rm x,1}}{n_{\rm x,total}}\times q_{\rm xk},\quad x=\romannumeral1-\romannumeral5,k=1-7.
	\label{eq:eta}
\end{equation}
$n_{\rm x,1}$ and $n_{\rm x,total}$ represent the number of observed asteroids with $15\leq H<16$ and $15\leq H<22$ in zone $x$, respectively. For a `near-gap' simulation (A--F), determining its associated zone and magnitude interval is sufficient to calculate $\eta$. For the `far-gap' simulation (G), it spans across multiple zones. Hence, starting from the outputs, we calculate the average value for zones belonged to each recorded CAPHA. Therefore, $\eta$ of Earth-CAPHAs and Mars-CAPHAs for simulation~G are different. Then if one simulation contributes $N$ CAPHAs, we correct the quantity to $\eta N$. The $\eta$ values of all simulations are shown in column~12 of Table~\ref{tab:setting_result}.

\begin{figure}
    \includegraphics[width=\columnwidth]{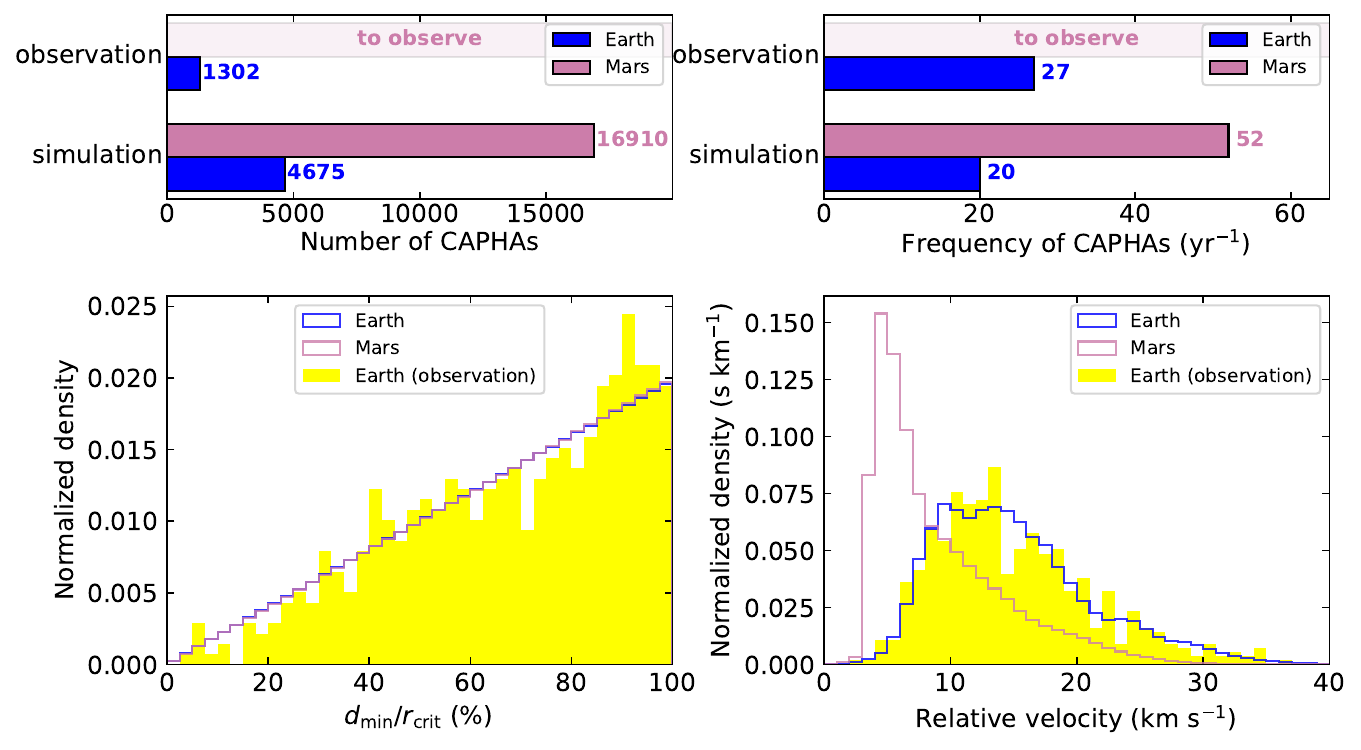}
    \caption{Upper left (right) panel: Numbers (Frequencies) of Earth-CAPHAs and Mars-CAPHAs obtained from simulations and actual observations. Lower left panel: Distribution of relative distances between CAPHAs and the planet. The relative distance of a CAPHA in percentage is defined as the ratio between its $r_{\rm min}$ and $r_{\rm crit}$ (0.05\,au for Earth-CAPHAs and 0.036\,au for Mars-CAPHAs). Blue and reddish purple lines almost coincide with each other. The yellow histogram shows the actually observed distribution of Earth-CAPHAs. Lower right panel: Distribution of velocities of CAPHAs relative to the planet. The yellow histogram shows the actually observed distribution of Earth-CAPHAs.}
    \label{fig:2}
\end{figure}

Lastly, we multiply the quantity $N^{*}$ by coefficients $\epsilon$ and $\eta$ and sum up results of all simulations to estimate the  observable number of CAPHAs:
\begin{equation}
    N_{\rm obs}=\epsilon_{\rm G}\eta_{\rm G}N^{*}_{\rm G}+\sum\limits_{X=A}^{F}\sum\limits_{k=1}^{7}\epsilon_{\rm Xk}\eta_{\rm Xk}N^{*}_{\rm Xk}.
	\label{eq:number}
\end{equation}
We obtain a total count of 46.75 Earth-CAPHAs and 169.10 Mars-CAPHAs. Considering that our sample represents only 1\% of the actual main belt, the predicted total numbers of observable Earth-CAPHAs and Mars-CAPHAs are 4675 and 16910, i.e., a ratio of approximately 1:3.6 (see Fig.~\ref{fig:2}).

Then we estimate their expected occurrence frequencies (in units of yr$^{-1}$). We first calculate the average frequency of each CAPHA for each simulation:
\begin{equation}
    \bar f_{\rm Xk}=\frac{1}{N_{\rm Xk}}\sum\limits_{m=1}^{N_{\rm Xk}}\frac{C_{\rm Xk,m}}{t_{\rm act,Xk,m}}.
	\label{eq:frequency_f}
\end{equation}
$N_{\rm Xk}$ represents the number of CAPHAs generated in simulation~Xk (e.g., A1), $C$ denotes the repeating occurrence count of a CAPHA, and $t_{\rm act}$ is its active lifetime. Next, for equation~(\ref{eq:number}), we multiply each term in the summation by its corresponding $\bar f$, yielding the total occurrence frequency:
\begin{equation}
    f_{\rm obs}=\epsilon_{\rm G}\eta_{\rm G}N^{*}_{\rm G}\bar f_{\rm G}+\sum\limits_{X=A}^{F}\sum\limits_{k=1}^{7}\epsilon_{\rm Xk}\eta_{\rm Xk}N^{*}_{\rm Xk}\bar f_{\rm Xk}.
	\label{eq:frequency_F}
\end{equation}

The resultant occurrence frequencies of Earth-CAPHAs and Mars-CAPHAs are 20 and 52 per year, respectively, making the latter approximately 2.6 times higher. It is noteworthy that the exact numbers and frequencies depend on the adopted intrinsic size distribution of the main belt asteroids given by equation~(\ref{eq:exp}), but the ratio of either quantity between Mars-CAPHAs and Earth-CAPHAs is insensitive to it.

Over 1300 independent (counted without repetition) Earth-CAPHAs have been observed and catalogued as of June 26, 2023, indicating that there are still at least 2.5 times more left to be discovered. The simulation predicted occurrence frequency of Earth-CAPHAs (20 per year) is in rough agreement with the observed frequency, which is about 27 per year averaged over 20 years, in view of the potential underestimation of the contribution from the `far-gap' region (Section~\ref{subsec:setting}). Both in terms of number and frequency, Mars-CAPHAs significantly surpass Earth-CAPHAs, indicating tremendous observational prospects.

It is worth noting that a Mars-CAPHA may become an Earth-CAPHA later, and vise verse. For example, in simulation~A2, a total of 72 Earth-CAPHAs and 125 Mars-CAPHAs were recorded, with 69 of them being in common. This raises an intriguing possibility of using CAPHAs as matter carriers between Earth and Mars.

\subsection{Distances and velocities}
We have further investigated the distance and velocity distributions of CAPHAs. Considering that the maximum distances $r_{\rm crit}$ from the central planet to Earth-CAPHAs and Mars-CAPHAs are not the same (0.05\,au for Earth and 0.036\,au for Mars, see section~\ref{sec:introduction}), we focus on the relative distance, which is defined as the ratio of a CAPHA's close-approach distance $r_{\rm min}$ and its corresponding planet's $r_{\rm crit}$. Fig.~\ref{fig:2} shows that Earth-CAPHAs and Mars-CAPHAs exhibit a similar distribution in terms of relative distances, and the actual observed results for Earth-CAPHAs align well with this distribution. Certainly, in terms of absolute distance, the majority of Mars-CAPHAs are closer to the planet than Earth-CAPHAs, making them more favorable for observations.

Moreover, the relative velocity (which remains nearly constant during a close approach of a given asteroid) of Mars-CAPHAs is systematically smaller compared to Earth-CAPHAs (see Fig.~\ref{fig:2}). It can be explained by the fact that, for the same orbit, when an asteroid is near Earth, it is closer to its perihelion, resulting in a higher velocity. This implies that the observing window is longer for Mars-based observations, or in other words, tracking CAPHAs from Mars is easier. The actually observed distribution of relative velocities for Earth-CAPHAs is also shown in Fig.~\ref{fig:2}, which matches our simulations well, demonstrating the validity of our simulations.

\subsection{Active lifetime of different CAPHAs}
The longer the active lifetime (defined in section~\ref{sec:results_quantity}) of a CAPHA, the higher the probability that it will coexist with the civilization on a particular planet, thereby increasing the chances of its observation. Our simulations reveal two main aspects of the active lifetime of CAPHAs.

First, the average active lifetime of Mars-CAPHAs (the longest reaching several tens of million years) is longer than that of Earth-CAPHAs (mostly ranges from several hundred thousand to several million years; see $\epsilon$ values in Table~\ref{tab:setting_result}), which can be explained by the difference in their eccentricities. Since Earth is farther from the main belt compared to Mars, an asteroid requires a higher eccentricity to become an Earth-CAPHA. A higher eccentricity implies a greater likelihood of close encounters with planets such as Jupiter \citep[e.g.][]{granvik2018}, which will kick asteroids out from the inner solar system, thus resulting in a shorter active lifetime.

Second, as the source region in the main belt becomes more distant, the average active lifetime of CAPHAs for a specific planet (Earth or Mars) decreases initially and then slightly increases, with a turning point occurring at around 2.96\,au (see Table~\ref{tab:setting_result}). Such a trend is determined by two factors: eccentricity and period. On one hand, the farther the source region is, the higher the eccentricity would be required for an asteroid to become a CAPHA, resulting in a shorter active lifetime. On the other hand, asteroids with larger semi-major axes have longer orbital periods, resulting in a lower frequency of close encounters with planets, and consequently, a longer active lifetime. For the nearer source regions, the first factor prevails, while as the distance increases, the second factor becomes more significant.

\subsection{Evaluating the feasibility of Earth-based observations}
In order to investigate the feasibility of Earth-based observations for Mars-CAPHAs, it is necessary to know their apparent magnitudes. Apparent magnitude $V(d,\phi)$ of an asteroid (or a planet) is determined by its absolute magnitude $H$, distance $d$ and phase angle $\phi$, and can be expressed simply as:
\begin{equation}
    V(d,\phi)=H+f(d,\phi).
	\label{eq:vh}
\end{equation}
Due to the close proximity of Mars-CAPHAs and Mars, their $d$ and $\phi$ can be taken as the same. As a result, they follow the same apparent magnitude conversion rules. 

The next Mars opposition (corresponding to $\phi=0$) will occur on January 16, 2025, so we focus on the time period around that date. Between December 10, 2024 and February 13, 2025, the apparent magnitude of Mars will be brighter than -0.52\,mag, according to the ephemeris. As the absolute magnitude of Mars is -1.52\,mag, the value of $f(d,\phi)$ during that period must be less than 1. Therefore, for Mars-CAPHAs appearing at that time, their apparent magnitude will not exceed their absolute magnitude plus one. Assuming the limiting magnitude of Earth-based observing facility is $V_{\rm lim}=20$\,mag, we need to consider Mars-CAPHAs with $H\leq 19$\,mag. Based on our simulations, the observable frequency of Mars-CAPHAs is at least 52 per year. According to equation~(\ref{eq:exp}), more than 22\% of Mars-CAPHAs have an absolute magnitude brighter than 19\,mag, which corresponds to over 11 occurrences per year. Therefore, during the aforementioned 65-day period, at least 2 Mars-CAPHAs can be observed from Earth. At that time, Earth-based telescopes, such as LSST \citep{ivezic2019} and WFST \citep{wang2023}, may have the best chance to detect Mars-CAPHAs, which can provide guidance for the planning of Mars-based observations.
\section{Summary}
\label{sec:summary}
We utilize the \textit{N}-body software package \textit{Mercury6} with an additional Yarkovsky effect subroutine to investigate Earth-CAPHAs and Mars-CAPHAs. Based on the assumption that CAPHAs originate from the main belt, we randomly sample 1\% of main belt asteroids as the initial conditions for our simulations. Considering special properties of gaps in the main belt, we divide our sample into `far-gap' asteroids and `near-gap' asteroids, and run them with different configurations.

The number and occurrence frequency of Mars-CAPHAs are a few times greater than that of Earth-CAPHAs, and the relative velocities of Mars-CAPHAs are favorable for Mars-based observations. Our results point to a new potential scientific objective for future Mars exploration missions (Mars~2020, Tianwen and so on). Observations of Mars-CAPHAs will deepen our understanding of the Martian environment, the interactions between asteroids and planets, and the evolutionary history of the inner Solar System. Lastly, a few Mars-CAPHAs are predicted to be observable even from Earth in 2025.

\section*{Acknowledgements}
We thank Tong Bao, Yun Su and Liangliang Yu for helpful discussions. We acknowledge a very helpful referee report. This work is supported by the National Natural Science Foundation of China (grants 12225302, 12373081 and 12150009).

%%%%%%%%%%%%%%%%%%%%%%%%%%%%%%%%%%%%%%%%%%%%%%%%%%
\section*{Data Availability}
The data underlying this article will be shared on reasonable request to the corresponding author.



\bibliographystyle{mnras}
\bibliography{martians} % if your bibtex file is called example.bib

%\begin{thebibliography}{99}
%\end{thebibliography}

%%%%%%%%%%%%%%%%%%%%%%%%%%%%%%%%%%%%%%%%%%%%%%%%%%

%%%%%%%%%%%%%%%%% APPENDICES %%%%%%%%%%%%%%%%%%%%%
%\appendix

% Don't change these lines
\bsp	% typesetting comment
\label{lastpage}
\end{document}